\documentclass[]{pasj01}
\usepackage[switch,mathlines]{lineno}

\begin{document} 

\Received{}
\Accepted{}

\title{Suzaku observation of an iron K-shell line in the spiral galaxy NGC 6946}

\author{Shigeo \textsc{Yamauchi}$^{\ast}$, \altaffilmark{} %
Azusa \textsc{Inaba}, 
and 
Yumiko \textsc{Anraku}}
\altaffiltext{}{Faculty of Science, Nara Women's University, Kitauoyanishimachi, Nara, Nara 630-8506, Japan}
\email{yamauchi@cc.nara-wu.ac.jp}

\KeyWords{galaxies: individual (NGC 6946) ---  X-rays: galaxies --- X-rays: ISM} 

\maketitle

\begin{abstract}
An emission line at $\sim$6.7 keV is attributable to a He-like iron K-shell transition, 
which indicates existence of a thin thermal plasma with a temperature of several keV. 
Using Suzaku archival data, we searched for the iron K-line from the spiral galaxy NGC 6946, and 
found the iron K-line at 6.68$\pm$0.07 keV at the 3.1$\sigma$ level in the central $r\le$\timeform{2.'5} region. 
The iron line luminosity from the central region was estimated to be (2.3$\pm$1.2)$\times$10$^{37}$ erg s$^{-1}$ at a distance of 5.5 Mpc. 
The origin of the iron emission line is discussed.
\end{abstract}


\section{Introduction}

Since the launch of the Einstein satellite with imaging capability, X-ray observations resolved individual X-ray sources 
and revealed the existence of diffuse X-ray emission with a temperature of $<$1 keV
(e.g., \cite{Fabbiano1989,Fabbiano1992,Read1997,Tyler2004,Owen2009,Mineo2012}).
The diffuse X-ray emission is thought to be X-rays from hot gas shock-heated by supernova (SN) explosions 
and stellar wind interactions.

An emission line at $\sim$6.7 keV is attributable to a K-transition line from He-like iron, 
which indicates existence of a thin thermal plasma with a temperature of several keV. 
The $\sim$6.7 keV emission line was detected from 
starburst galaxies (e.g., \cite{Pietsch2001,Boller2003,Strickland2007,Ranalli2008,Mitsuishi2011}) and ultraluminous infrared galaxies 
(e.g., Iwasawa et al. 2005, 2009). 
The high temperature plasma was thought to be produced by multiple SNe
after past starburst activities.

NGC 6946 is a face-on galaxy with four spiral arms. 
Its distance is estimated to be 5.5 Mpc \citep{Tully1988} -- 7 Mpc \citep{Israel1980}. 
NGC 6946 has strong nuclear starburst activity (e.g., \cite{Telesco1980,Elmegreen1998}), whereas 
further evidence for the high star forming activity, especially in the northern spiral arm, is indicated 
by CO radio emission line observations (e.g., \cite{Nieten1999,Walsh2002}). 
The total star formation rate is estimated to be 3.2 $M_{\odot}$ yr$^{-1}$ \citep{Jarrett2013}.
Active star formation in the past leads many SN explosions.  
In fact, 10 historical SNe have been observed since 1917 and 
a recent optical search for supernova remnants (SNRs) found many possible SNRs \citep{Long2019}. 

\begin{figure*}
  \begin{center}
        \includegraphics[width=14cm]{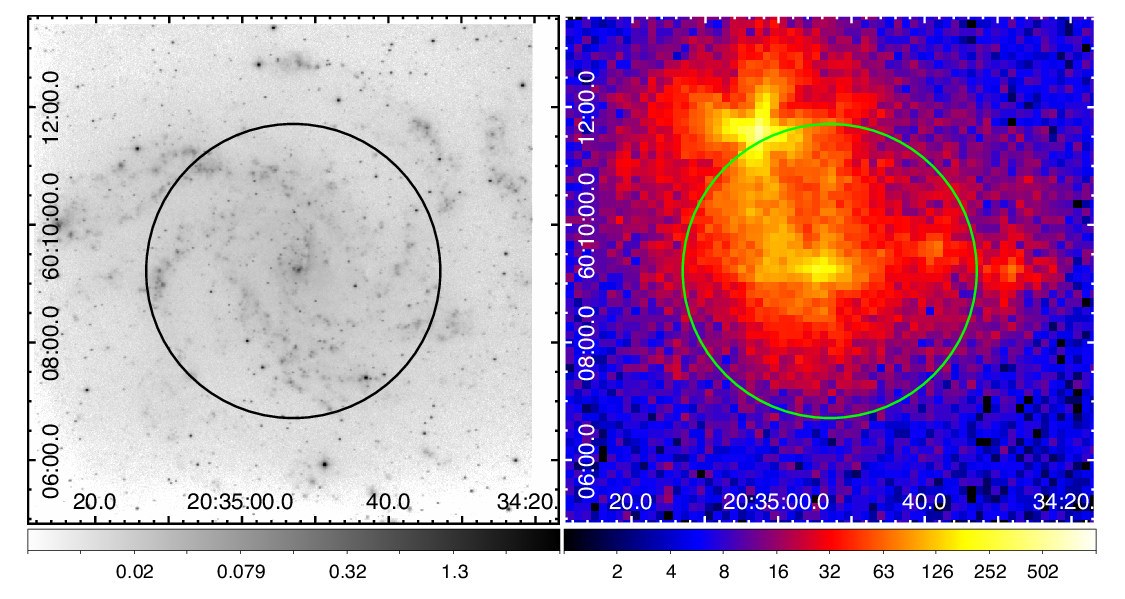}
  \end{center}
\caption{NGC 6946 H$\alpha$ image taken from NASA/IPAC Extragalactic Database (https://ned.ipac.caltech.edu) (left) and 
XIS image in the 1--8 keV energy band (right). 
The coordinates are J2000.0.  
The intensity levels are logarithmically spaced.
Data obtained with XIS 0, 1, and 3 sensors are merged.
A solid circle shows a source region from which an X-ray spectrum was extracted.
 }\label{fig:img}
\end{figure*}

X-ray observations of NGC 6946 have been carried out with Einstein, ROSAT, Chandra, and XMM-Newton 
and have revealed properties of X-ray emissions.  
In addition to discrete X-ray sources, a weak diffuse emission was detected (e.g., \cite{Schlegel1994,Schlegel2003,Wezgowiec2016}). 
\citet{Schlegel2003} estimated that as much as 10\% of the total soft X-ray emission could be due to a diffuse component and showed that 
the diffuse component was explained by a two-component thermal plasma model with temperatures of 0.25$\pm$0.03 keV and 0.70$\pm$0.10 keV. 
Based on the XMM-Newon observation, \citet{Wezgowiec2016} revealed that parameters of the diffuse plasma exhibit spatial variation:  
two-temperature plasma models are needed for the disk and halo emission 
($kT_{\rm e}\sim$0.2--0.3 keV and 0.6--0.8 keV), while the inter-arm regions show only one thermal component ($kT_{\rm e}\sim$0.4--0.6 keV). 
 \citet{Wezgowiec2016} proposed that an increase of the temperature in the magnetic arm regions could be described as additional heating due to 
 magnetic reconnection. 
 
Is there a high temperature plasma having a 6.7 keV line in NGC 6946? 
 \citet{Wezgowiec2016} did not state the existence of the iron K-line in the spectra. 
The detection of the iron K-line has not been reported so far. 
According to \citet{Snowden2008}, 
the non X-ray background (NXB) of XMM-Newton is modeled as a power law function representing the soft proton contamination 
plus Gaussians representing instrumental fluorescence lines. 
The soft proton component has a hard spectrum extending  up to $\sim$12 keV and is time-variable, 
which would affect estimation of a continuum shape and the iron K-line intensity of the diffuse component. 

The X-ray Imaging Spectrometers (XIS, \cite{Koyama2007a}) onboard the Suzaku satellite \citep{Mitsuda2007} 
has sufficient energy resolution and lower/more stable NXB than the previous X-ray satellites, and hence 
the XIS has the best sensitivity in the iron K-line band. 
Suzaku observed NGC 6946 with a long exposure time in a single observation ($\sim$200 ks), rather than several occasions as XMM-Newton. 
Therefore, the Suzaku XIS data is suitable to search for the iron K-line. 
In this paper, we present the results of a spectral analysis of NGC 6946 using the Suzaku XIS data and discuss the origin of the iron emission line. 
All through this paper, the errors are given as 90\% confidence level.

\section{Observation and date reduction}

X-ray observation of NGC 6946 was made on 2012 May 13--16 with the Suzaku XIS \citep{Koyama2007a},
which was placed on the focal planes of the thin foil X-ray Telescopes (XRT, \cite{Serlemitsos2007}).  
The field of view (FOV) of the XIS was \timeform{17.'8}$\times$\timeform{17.'8}.
The XIS was composed of 4 sensors: 
XIS sensor-1 (XIS 1) was a back-side illuminated CCD (BI), while
the other three XIS sensors (XIS 0, 2, and 3) were front-side illuminated CCDs (FIs).
However, one of the FIs  (XIS 2) stopped working on 2006 November 9.
Therefore, we utilized data obtained with three sensors, XIS 0, XIS 1, and XIS 3 for the following analysis.
The XIS was operated in the normal clocking mode.

The Suzaku archival data were retrieved from the DARTS system operated by ISAS/JAXA.
Data reduction and analysis were made using the HEAsoft version 6.30.1
and the calibration database version 2018-10-10. 
The exposure time was 198.9 ks. 
In addition to the standard screening, we processed noisy pixel rejection  
according to the web page\footnote{https://heasarc.gsfc.nasa.gov/docs/suzaku/analysis/xisnxbnew.html}.  

\section{Analysis and results}

Figure \ref{fig:img} shows an H$\alpha$ image 
and an XIS image in the 1--8 keV energy band.
Several point-like X-ray sources are seen.  
The most luminous source located at the northeast from the center  
is identified with an ultra-luminous X-ray source, NGC 6946 X-1 (e.g., \cite{Fabbiano1987}), 
associated with the optical nebula MF16 \citep{Matonick1997}. 
NGC 6946 X-1 exhibits a featureless spectrum (e.g., \cite{Roberts2003,Ghosh2023}), and it is thought to be a black hole binary. 
Results of a spectral analysis for NGC 6946 X-1 are presented in Appendix. 
In addition to the discrete sources, there seems to be a faint extended emission as is reported by \citet{Wezgowiec2016}. 

In order to examine existence of an iron K-line in NGC 6946, we carried out a spectral analysis. 
X-ray spectrum was extracted from a circle with a radius of \timeform{2.'5} (source region, a circle in figure \ref{fig:img}).
The NXB was taken from the night earth data 
using {\tt xisnxbgen} \citep{Tawa2008}.
After subtracting the NXB, we merged the FI spectra to maximize photon statistics.
Response files, Redistribution Matrix Files (RMFs) and Ancillary Response Files (ARFs), 
were made using {\tt xisrmfgen} and {\tt xissimarfgen} \citep{Ishisaki2007}, respectively. 
In making the ARF, we assumed a uniform sky. 

In order to maximize photon statistics, we added data with XIS 0 and 3 and grouped the data with a minimum of 40 counts per bin for XIS 0+3 and 
20 counts per bin for XIS 1 after subtracting the NXB. 
The spatial resolution of the Suzaku XIS is worse than those of Chandra and XMM-Newton, and hence the source spectrum contains X-rays 
from not only diffuse X-ray emission but also discrete sources. 
In order to focus on searching for the iron K-line, we utilized the iron K-line band data.
 
\begin{figure}[t]
  \begin{center}
      \includegraphics[width=8cm]{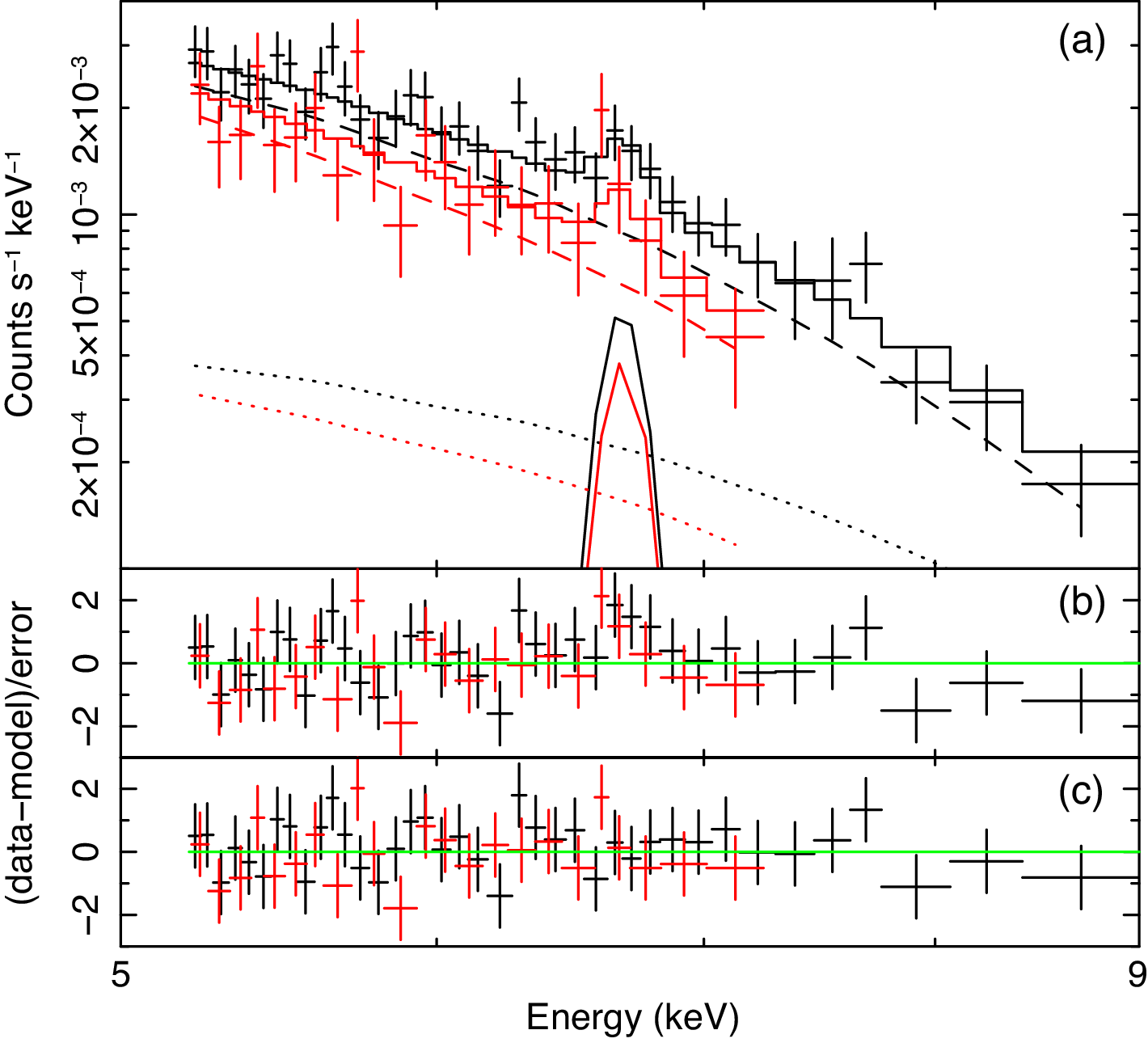}
        \end{center}
  \caption{
(a) XIS spectra of the central region of NGC 6946 (black: XIS 0+3 and red: XIS 1) and the best-fitting model (histogram). 
The dashed, dotted, and the solid lines show contributions of  
continuum emission (thermal bremsstrahlung), the CXB, and an iron K-line, respectively. 
(b) Residuals from the best-fitting model without the iron K-line. 
(c) Residuals from the best-fitting model with the iron K-line.  
}\label{fig:spc1}
\end{figure}

Figure \ref{fig:spc1}a shows the NXB-subtracted spectra in the 5--9 keV energy band. 
At first, we fitted the spectra with a bremsstrahlung + the Cosmic X-ray background (CXB) model modified by a low-energy absorption. 
The spectral parameters of the CXB were fixed to the values in \citet{Kushino2002}, 
while the cross sections of the absorption were taken from \citet{Verner1996}. 
As a result, we found positive residuals around 6.7 keV in the spectrum, as is shown in figure \ref{fig:spc1}b.
We added an emission line model with a line width of null to the spectra and found that the fit was improved (figure \ref{fig:spc1}c). 
The $\Delta {\chi}^2$ value was 10, which shows the F-test probability of 0.002. 
The significance level estimated from the obtained photon flux and its statistical error is 3.1$\sigma$.  
The best-fitting parameters are listed in table \ref{tab:spc1} and the best-fitting model is plotted in figure \ref{fig:spc1}a. 
After removing contribution of NGC 6946 X-1, an equivalent width was estimated to be 0.19$\pm$0.10 keV. 

The 6.7 keV emission line originates from a hot plasma with a temperature of several keV. 
We also examine the existence of 6.97 keV line attributable to a H-like iron K-shell transition. 
However, we found no obvious emission line and gave an upper limit of 4.0$\times$10$^{-7}$ photons s$^{-1}$ cm$^{-2}$ 
(90\% confidence level).

%
\begin{table}[t]
\caption{The best-fitting parameters of the central region.}
\begin{center}
\begin{tabular}{lc} \hline  
Parameter & Value\\
\hline 
\multicolumn{2}{c}{Model: phabs$\times$(bremsstrahlung+line+CXB)}\\
$N_{\rm H}$ (cm$^{-2}$) 			& 2.2$\times$10$^{21}$$^{\ast}$ (fixed) \\
$kT_{\rm e}$ (keV)   				&  3.4$^{+1.2}_{-0.7}$ \\
$E_{\rm Line}$ (keV) 			& 6.68$\pm$0.07 \\
$I_{\rm Line}$ (photons s$^{-1}$ cm$^{-2}$) & (5.9$\pm$3.1)$\times$10$^{-7}$ \\
\hline
Photon index$_{\rm CXB}$ 		& 1.412 (fixed)\\
Norm$_{\rm CXB}^{\dag}$ 	& 8.17$\times$10$^{-7}$ (fixed)\\
$\chi^2$/d.o.f. & 42.4/58    \\
\hline \\
\end{tabular}
\end{center}
\vspace{-12pt}
$^{\ast}$ \citet{HI4PI2016}.\\
$^{\dag}$ In units of photons s$^{-1}$ cm$^{-2}$ arcmin$^{-2}$ at 1 keV. \\
\label{tab:spc1}
\end{table}

\section{Discussion}

Using the Suzaku data, we detected an iron K-line at 6.68$\pm$0.07 keV with a 3.1$\sigma$ level. 
The previous work using XMM-Newton \citep{Wezgowiec2016} did not report 
whether the iron K-line was found in the spectra or not.  
Thus, we analyzed the XMM-Newton archival data (7 occasions, total accumulated time=229.9, 232.7, and 177.9 ks for MOS1, MOS2, and pn, respectively) 
and examined whether the iron K-line can be seen. 
We fitted the spectra extracted from a central $r\le$\timeform{2.'5} region (same as that of the Suzaku data analysis) 
with the same spectral model as that in the Suzaku observation  
and found a hint of  the iron K-line at 6.68 keV. 
The line flux was estimated to be (7.3$\pm$3.0)$\times$10$^{-7}$ photons s$^{-1}$ cm$^{-2}$ (4.0 $\sigma$ level), consistent with that in the Suzaku observation. 

Using the best-fitting parameters for the Suzaku spectrum, we calculated the iron line luminosity to be $L_{\rm Fe}$=(2.3$\pm$1.2)$\times$10$^{37}$ erg s$^{-1}$  
within a \timeform{2.'5} radius, a 4 kpc radius at a distance of 5.5 Mpc \citep{Tully1988}.  
\citet{Schlegel2003} and \citet{Wezgowiec2016} reported the existence of a hot plasma with a temperature of $\sim$0.7--0.8 keV. 
The iron line centroid is less than 6.6 keV in the 0.7--0.8 keV plasma. 
Since the lower end of the observed line centroid is 6.61 keV, 
we can safely conclude that NGC 6946 has a hotter plasma than those found by \citet{Schlegel2003} and \citet{Wezgowiec2016}. 
Following argument in M101 paper \citep{Yamauchi2016}, we discuss the origin of the iron emission line.


Stellar sources such as cataclysmic variables (CVs), active binaries (ABs), and young stellar objects (YSOs) in star forming regions 
are known to have a thin thermal emission with the iron emission line 
(e.g., \cite{Ezuka1999,Gudel1999,Yamauchi1996}).
At first, we discuss the stellar source origin.

\citet{Sazonov2006} showed a cumulative luminosity density of the stellar sources in the Galaxy to be 
$L_{\star, }$$_{\rm 2-10 keV}$/$M_{\star}$=(4.5$\pm$0.9)$\times$10$^{27}$ erg s$^{-1}$ $M_{\odot}^{-1}$, 
where $L_{\star, }$$_{\rm 2-10 keV}$ is a stellar source luminosity in the 2--10 keV band and $M_{\star}$ is a stellar mass. 
Assuming the typical spectral model for CVs, ABs, and YSOs (e.g., \cite{Baskill2005,Ishida2009,Gudel1999,Yamauchi1996}), 
we can estimate a stellar source iron line luminosity, $L_{\star, }$$_{\rm Fe}$, to be 3--4\% of $L_{\star, }$$_{\rm 2-10 keV}$.   
The rotation velocity of NGC 6946 of $\sim$170 km s$^{-1}$ at the radius of 4 kpc \citep{deBlok2008} gives
the total mass within the central 4 kpc radius region of 2.7$\times$10$^{10}$$M_{\odot}$. 
This value is an upper limit of the stellar mass 
because galaxies contains interstellar matter, high mass stars, X-ray sources such as black hole binaries 
and neutron-star binaries, a supermassive black hole at the center, and dark mater as well as CVs, ABs, and YSOs. 
Assuming all the mass is equal to the total mass of CVs, ABs, and YSOs,  
we obtain $L_{\star, }$$_{\rm 2-10 keV}$=(0.9--1.5)$\times$10$^{38}$ erg s$^{-1}$, 
and then $L_{\star, }$$_{\rm Fe}$$\le 6 \times$10$^{36}$ erg s$^{-1}$.
Thus, the well-known stellar sources cannot account for all the observed $L_{\rm Fe}$. 
Taking account of contribution of the stellar sources, 
we can conclude that $L_{\rm Fe}$$\sim$10$^{37}$ erg s$^{-1}$  
originates from other objects or process.  
 

Young and middle-aged SNRs have a several keV temperature plasma and exhibit an intense iron K line. 
Thus, hot plasmas produced by SN explosions are a possible candidate.
The typical SNR luminosity of $L_{\rm SNR, }$$_{\rm 2-10keV}\sim$10$^{34-36}$ erg s$^{-1}$ 
(e.g., \cite{Seward2000}).  
Assuming a several keV temperature and solar abundances \citep{Anders1989}, 
we can estimate $L_{\rm SNR, }$$_{\rm Fe}$ to be $\sim$7\% of $L_{\rm SNR, }$$_{\rm 2-10keV}$, 
and hence $L_{\rm SNR, }$$_{\rm Fe}$ of each SNR to be $\sim$10$^{33-35}$erg s$^{-1}$. 
The total number of SNRs, $N_{\rm SNR}$ to account for the $L_{\rm Fe}$$\sim$10$^{37}$ erg s$^{-1}$ is $\sim$100--10$^4$.
According to the Sedov solution \citep{Sedov1959}, 
a typical SNR age with a temperature of $>$1 keV ($t$) is $\le$4$\times$10$^{3}$ yr. 
Thus, a SN rate to explain the observed $L_{\rm Fe}$ is $N_{\rm SNR}$/$t$$\sim$0.03--3 yr$^{-1}$ within the 4 kpc radius. 
The SN rate is estimated to be $\sim$0.1 yr$^{-1}$ based on the fact that 10 SNe have been discovered since 1917. 
Multiple SN explosions following the active star formation can produce a large amount of diffuse hot plasma. 

Since many SNRs and SNR candidates were discovered in the whole galaxy \citep{Long2019}, 
SNe in the Galactic arm would contribute to producing the diffuse hot plasma considerably. 
The image obtained with XMM-Newton showed a diffuse emission around the central region of NGC 6946 
as well as along the Galactic arm \citep{Wezgowiec2016}. 
The 6.7 keV line enhancement was found in the central region of the starburst galaxy NGC 253 \citep{Mitsuishi2011}. 
Substantial iron line emission may associate with the activity in the central region such as a nuclear starburst. 

\citet{Schlegel2003} and \citet{Wezgowiec2016} reported existence of a hot plasma with a temperature of $\sim$0.7--0.8 keV in NGC 6946 
and slightly enhancement of the temperature in the arm regions. 
\citet{Wezgowiec2016} estimated the magnetic field strength to be $\sim$10--20 $\mu$G and proposed magnetic reconnection as additional heating. 
 As a scenario of the Galactic Ridge X-ray Emission (GRXE), which is distributed along the Galactic plane and exhibits an intense iron K-line (e.g., \cite{Koyama2018}), 
 \citet{Tanuma1999} showed that the magnetic reconnection heats a cool ($\sim$0.8 keV) plasma generated by SNe up to the hot ($\sim$7 keV) plasma, 
 if the local magnetic field strength is $\sim$30 $\mu$G with a filling factor of $<$0.1. 
Here, we discuss on possibility of the magnetic activity based on the present results. 

Adopting a temperature of 3 keV from the spectral fitting (see table 1) 
and solar abundance for the hot plasma and $L_{\rm Fe}$$\sim$10$^{37}$ erg s$^{-1}$ within a \timeform{2.'5} radius, 
we estimated the luminosity in the 2--10 keV band to be $\sim$10$^{38}$ erg s$^{-1}$.  
Assuming the X-ray emitting volume of a cylinder with a radius of 4 kpc and a height of 400 pc and $n_{\rm e}=1.2 n_{\rm H}$, 
where $n_{\rm e}$ and $n_{\rm H}$ are electron and hydrogen densities, respectively, 
we obtained $n_{\rm H}$ of $\sim$5$\times$10$^{-3} f^{-0.5}$ cm$^{-3}$ and a thermal energy density of  
$\sim$7$\times$10$^{-11}f^{-0.5}$ erg cm$^{-3}$, where $f$ is a filling factor. 
Since the total thermal energy is larger than the energy density of magnetic field of (0.7--1.9)$\times$10$^{-11}$ erg cm$^{-3}$ \citep{Wezgowiec2016}, 
it is difficult to account for all the hot plasma by only the magnetic energy.  

Using XMM-Newton data of M101, \citet{Wezgowiec2022} argued what amount of the magnetic energy density would be converted to thermal energy 
by the magnetic reconnection and consequently increase the temperature of the hot gas. 
In their estimation, a magnetic energy density of $\sim$2$\times$10$^{-13}$ erg cm$^{-3}$, which is about 10\% of the total magnetic energy density, 
would be converted to additional thermal energy of $\sim$10$^{-10}$ erg per particle in the hot gas having a density of $\sim$10$^{-3}$ cm$^{-3}$,  
which leads to an increase in temperature by around 0.1 keV. 
Applying the same estimation to the case of NGC 6946, we found that 
the total magnetic energy density of $\sim$(1--2)$\times$10$^{-11}$ erg cm$^{-3}$ is enough to add 
a thermal energy of $\sim$10$^{-10}$ erg per particle in the hot gas.
Thus, additional heating for primary hot plasma, probably produced by SNe, by the magnetic reconnection would be possible. 

A hot plasma with a temperature of a few keV cannot be bound by the galactic gravity and it would cool down once the reconnection stops to operate.  
Therefore, if this process works, continuous heating by reconnection and some processes to maintain the hot plasma 
such as magnetic confinement proposed by \citet{Makishima1994} should be required. 
To investigate this process in detail, we need to know spatial distribution and physical parameters (temperature, density, thermal energy, and so on) 
of the hot plasma after removing contribution of point sources. 

\citet{Masai2002} proposed a scenario of the origin of the GRXE based on stochastic particle acceleration in the interstellar medium. 
Free electrons in the hot interstellar medium produced by SNe are scattered and accelerated through a stochastic process 
by irregular magnetic fields left behind SNR shocks.
The electron spectrum is reproduced approximately by the sum of the a few hundred eV Maxwellian of thermal electrons produced by SNe, 
power-law non-thermal component, and a few keV Maxwellian of quasi-thermal electrons.  
A thermal plasma at a few hundred eV can be bound by the galactic gravity. 
Highly ionized iron is produced by ionization by the quasi-thermal electrons. 
In addition, a fraction of ions also captures background thermal electrons, i.e., recombining. 
Thus, in the case of this model, the iron K-line and many recombination edges on the non-equilibrium continuum spectrum, 
especially recombination structure by iron ion at 8--9 keV, can be seen in the spectrum.   
However, we see no recombination structure in the Suzaku spectrum due to limited photon statistics. 

\section{Conclusion}

Using Suzaku archival data, we detected the iron K-line at 6.68 keV at the 3.1$\sigma$ level in the central $r\le$\timeform{2.'5} region of NGC 6946 for the first time. 
The iron line luminosity from the central region was estimated to be (2.3$\pm$1.2)$\times$10$^{37}$ erg s$^{-1}$ at a distance of 5.5 Mpc. 
Contribution of stellar sources in NGC 6946 is not enough to explain the total iron K-line luminosity, and hence
$L_{\rm Fe}$$\sim$10$^{37}$ erg s$^{-1}$ originates from other objects or process. 
SNe with a high SN rate are considered to be a possible scenario. 
The magnetic energy density is not enough to account for all the hot plasma with the iron K-line, 
but additional heating for primary hot plasma by the magnetic reconnection would be possible.
A quasi-thermal electrons produced by a stochastic acceleration may be the origin of the iron K-line. 
Unfortunately, due to limited spatial resolution and photon statistics of the Suzaku observation, 
we cannot conclude which process works. 
We need further observations with high spatial resolution and good photon statistics to reveal the origin of the emission line. 

\section*{Acknowledgements}

The authors are grateful to all members of the Suzaku team. 
The authors wish to thank the referee for constructive comments that improved the manuscript. 
This work was supported by the Japan Society for the Promotion of Science (JSPS) KAKENHI Grant Numbers JP 21K03615 (SY).

\section*{Appendix: X-ray spectrum of NGC 6946 X-1}

\begin{figure}[t]
  \begin{center}
      \includegraphics[width=8cm]{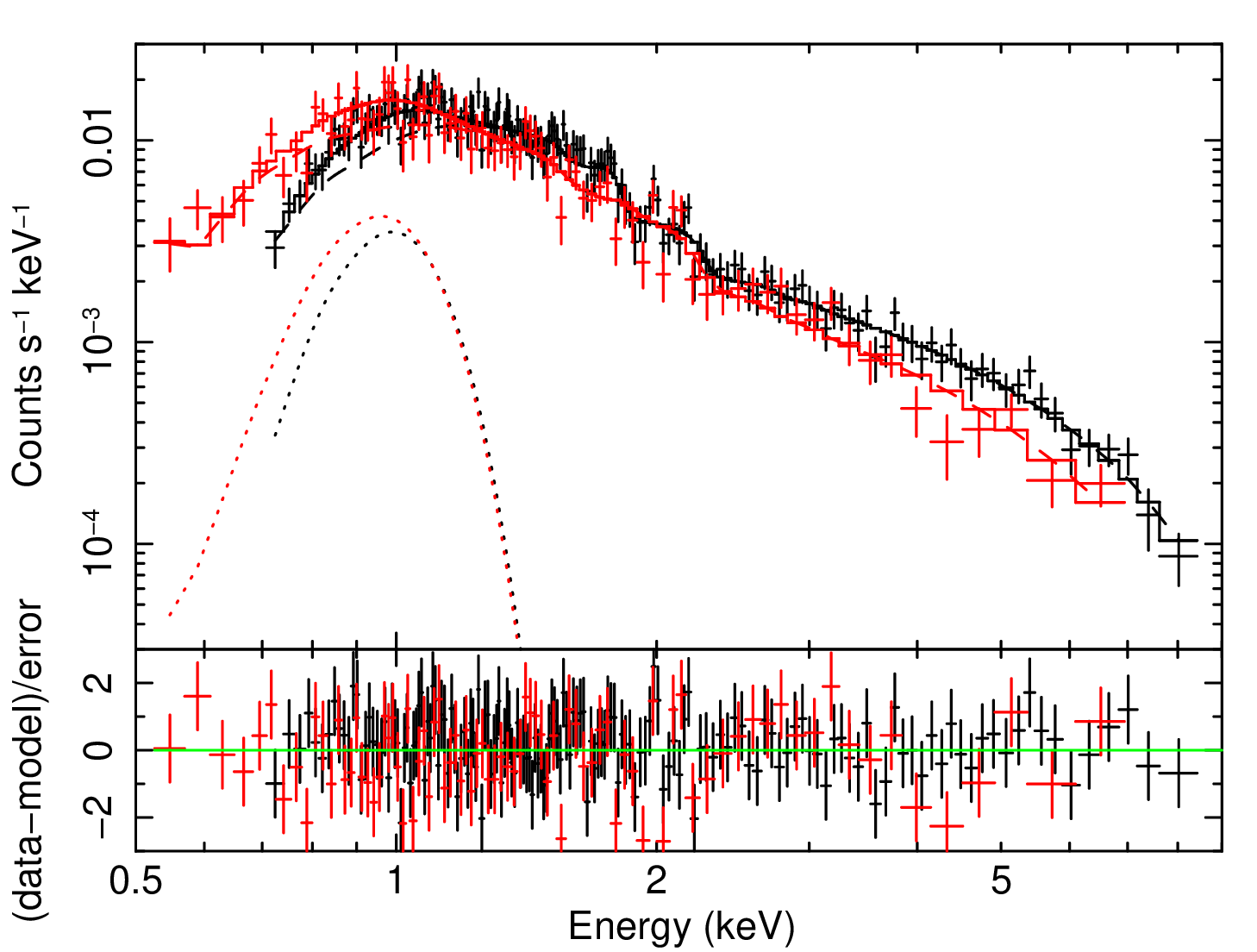}
        \end{center}
  \caption{
Upper panel: XIS spectra of NGC 6946 X-1 (black: XIS 0+3 and red: XIS 1) and the best-fitting model (histogram). 
The dashed and dotted lines show continuum emission and a broad Gaussian, respectively. 
Lower panel: Residuals from the best-fitting model.  
}\label{fig:spc2}
\end{figure}

The source spectrum was extracted from a 40$''$ radius circle. 
The background spectrum 
was extracted from an annulus with inner and outer radii of 60$''$ and 90$''$, respectively, but excluding the source region. 
We added data with XIS 0 and 3 and grouped the data with a minimum of 50 counts per bin for both XIS 0+3 and XIS 1 spectra. 
The ARF was made as a point source at the position of NGC 6946 X-1. 
The background-subtracted source spectra are shown in figure \ref{fig:spc2}.
We fitted the spectrum with model consisting of a multi-color disk ({\tt diskbb} model in XSPEC, \cite{Mitsuda1984}) 
and a thermally Comptonized continuum ({\tt thcomp} model in XSPEC, \cite{Zdziarski2020}) 
modified by a low energy absorption ({\tt phabs} model in XSPEC)
but found that the model showed a systematic broad residuals around 1 keV. 
Referring \citet{Ghosh2023}, we added a broad Gaussian at 0.9 keV and obtained an improved fit. 
The best-fitting parameters are listed in table \ref{tab:spc2}, while the best-fitting model is plotted in figure \ref{fig:spc2}. 
The luminosity was estimated to be $\sim$2$\times$10$^{39}$ erg s$^{-1}$ in the 2--10 keV band at 5.5 Mpc. 
We found no significant line at 6.7 keV ($<7\times10^{-7}$ photons s$^{-1}$ cm$^{-2}$, 90 \% confidence level). 

%
\begin{table}[t]
\caption{The best-fitting parameters of NGC 6946 X-1.}
\begin{center}
\begin{tabular}{lc} \hline  
Parameter & Value\\
\hline 
\multicolumn{2}{c}{Model: phabs$\times$(thcomp$\times$diskbb+gaussian)}\\ 
$N_{\rm H}$ (cm$^{-2}$) 							& 2.2$\times$10$^{21}$$^{\ast}$ (fixed) \\
$kT_{\rm in}$ (keV)								& 0.23$^{+0.08}_{-0.05}$\\
Norm										& 19$^{+34}_{-15}$ \\
Photon index									&  2.3$\pm$0.2 \\
$kT_{\rm e}$ (keV)								& 100 (fixed)\\
Covering fraction								& $>0.53$ \\
$E_{\rm Gaussian}$ (keV) 						& 0.90$^{+0.07}_{-0.12}$ \\
$\sigma_{\rm Gaussian}$ (keV)						& 0.14$^{+0.07}_{-0.05}$\\
$I_{\rm Gaussian}$ (photons s$^{-1}$ cm$^{-2}$) 		& (6.0$^{+9.8}_{-3.0}$)$\times$10$^{-5}$ \\
$\chi^2$/d.o.f. & 236.1/239    \\
\hline \\
\end{tabular}
\end{center}
\vspace{-12pt}
$^{\ast}$ \citet{HI4PI2016}.\\
\label{tab:spc2}
\end{table}


\end{document}